\documentstyle[12pt]{article}
\begin{document}
\title{Support for the Jaffe-Wilczek \\ Diquark Model of Pentaquarks}
\author{Rainer W. K\"uhne \\ 
{\it Vorm Holz 4, 42119 Wuppertal, Germany} \\
{\it kuehne70@gmx.de}}
\maketitle
\vskip 1cm
\begin{abstract}
\noindent
I examine the diquark model of pentaquarks that was suggested by Jaffe 
and Wilczek. Based upon this model, I predict the states $\Theta$(1530), 
N(1710), $\Sigma$(1880) and $\Xi$(1770) to be members of the same 
anti-decuplet. Moreover I predict the states N(1440), $\Lambda$(1600), 
$\Sigma$(1660) and $\Xi$(1950) to be members of the corresponding octet.
\end{abstract}
\vskip 1cm

\noindent
PACS: 12.38.Lg, 12.39.Mk, 12.40.Yx

\vskip 1cm

\noindent
Keywords: pentaquark, diquarks

\vskip 1cm

\noindent
Diakonov et al. \cite{1} predicted a pentaquark with the quark content 
$uudd\bar s$. They predicted its spin to be $J=1/2$, its parity to be 
positive, 
its isospin to be $I=0$, its hypercharge to be $Y=2$, its strangeness 
to be $S=+1$, its electric charge to be $Q=+e$, its mass to be 
$M=1530$MeV, and its width to be $\Gamma\le 15$MeV. This pentaquark is 
now named $\Theta^{+}$. 

The $\Theta^{+}$ has been detected by the LEPS Collaboration \cite{2} and 
confirmed by the DIANA Collaboration \cite{3}, the CLAS Collaboration 
\cite{4,5}, the SAPHIR Collaboration \cite{6}, the HERMES Collaboration 
\cite{7}, the SVD Collaboration \cite{8}, the COSY-TOF Collaboration 
\cite{9}, and by Asratyan et al. \cite{10} who examined data of the 
neutrino experiments WA21, WA25, WA59, E180, and E632. These experiments 
have confirmed the properties of the $\Theta^{+}$ to be $I_3 =0$, $Y=2$, 
$S=+1$, $Q=+e$, $M=(1526\ldots 1555)$MeV, and $\Gamma\le 20$MeV.

Its parity has not yet been determined experimentally. The simple 
quark model and the diamond structure model of the pentaquark predict 
its parity to be negative \cite{11,12}. However, Diakonov et al. \cite{1} 
predicted the parity of the $\Theta^{+}$ to be positive, because the 
parity of the N(1710), which should be a member of the same anti-decuplet, 
has positive parity.

Jaffe and Wilczek \cite{13} suggested that the pentaquark consists of 
two diquarks which are bound by the antiquark. They suggested a mass 
formula for the pentaquarks. The mass $M_0 = 1440$MeV is given by the 
mass of the lightest pentaquark which is assumed to be the Roper N(1440). 
Pentaquarks which include a strange quark instead of an up or down quark 
obtain an additional mass $m_s = 95$MeV for each strange quark. Diquarks 
which contain a strange quark are supposed to be less bound than diquarks 
which contain only up and down quarks. Pentaquarks which contain diquarks 
which include at least one strange quark obtain an additional mass 
$m_{\alpha} = 60$MeV for each diquark.

Thus, one can predict the masses of the pentaquarks according to the table 1.

\begin{table}
\caption{Properties of Pentaquarks: Anti-Decuplet and Octet}
\begin{center}
\begin{tabular}{ccrc}
\hline
Particle & Diquark content & Predicted mass & Observed mass \\
 & (example) & (MeV) & (MeV) \\
\hline
$\Theta$ & [$ud$]~[$ud$]$\bar s$ & $M_0 + m_s = 1535$ & 1530 \\
N & [$us$]~[$ud$]$\bar s$ & $M_0 + 2m_s + m_{\alpha}= 1690$ & 1710 \\
$\Sigma$ & [$us$]~[$us$]$\bar s$ & $M_0 + 3m_s + 2m_{\alpha}= 1845$ & 1880 \\
$\Xi$ & [$us$]~[$us$]$\bar d$ & $M_0 + 2m_s + 2m_{\alpha}= 1750$ & 1770 \\
\hline
N & [$ud$]~[$ud$]$\bar d$ & $M_0 = 1440$ & 1440 \\
$\Lambda$ & [$ds$]~[$ud$]$\bar d$ & $M_0 + m_s + m_{\alpha}= 1595$ & 1600 \\
$\Sigma$ & [$us$]~[$ud$]$\bar d$ & $M_0 + m_s + m_{\alpha}= 1595$ & 1660 \\
$\Xi$ & [$ss$]~[$us$]$\bar s$ & $M_0 + 4m_s + 2m_{\alpha}= 1940$ & 1950 \\
\hline
\end{tabular}
\end{center}
\end{table}
The $\Theta$(1530), N(1440), N(1710), $\Lambda$(1600), $\Sigma$(1660) , 
$\Sigma$(1880) and $\Xi$(1950) are 
well established particles. Recently, the CLAS Collaboration \cite{14} 
observed a cascade of $\Xi^{-}$ with masses 1321, 1530, 1620, 1690, 
1770, 1820, 1860, 1950, and 2030 MeV. The $\Xi$(1770) and $\Xi$(1860) 
are new states.

The $\Xi^{--}$(1862) reported by the NA49 Collaboration \cite{15} does 
not appear to be a member of the anti-decuplet considered in this paper. 

To conclude, this model predicts the existence of the $\Xi^{--}$(1770) 
and the $\Xi^{+}$(1770). Furthermore, it predicts the $\Theta$(1530), 
$\Xi$(1770) and $\Xi$(1950) to have $J^{P}= {\frac{1}{2}}^{+}$.

\end{document}